\documentclass{article}
\usepackage[utf8]{inputenc}
\usepackage[margin=1in]{geometry}

\title{HyperNetX: A Python package for modeling complex network data as hypergraphs}
\author{Brenda Praggastis 
    \and Sinan Aksoy 
    \and Dustin Arendt 
    \and Mark Bonicillo
    \and Cliff Joslyn
    \and Emilie Purvine
    \and Madelyn Shapiro
    \and Ji Young Yun}

\date{June 2023}

\usepackage{graphicx}

\begin{document}

\maketitle
 
\abstract{
HyperNetX (HNX) is an open source Python library for the analysis and visualization of complex network data modeled as hypergraphs.
Initially released in 2019, HNX facilitates exploratory data analysis of complex networks using algebraic topology, combinatorics, and generalized hypergraph and graph theoretical methods on structured data inputs.
With its 2023 release, the library supports attaching metadata, numerical and categorical, to nodes (vertices) and  hyperedges, as well as to node-hyperedge pairings (incidences).
HNX has a customizable Matplotlib-based visualization module as well as HypernetX-Widget, its JavaScript addon for interactive exploration and visualization of hypergraphs within Jupyter Notebooks. Both packages are available on GitHub and PyPI. With a growing community of users and collaborators, HNX has become a preeminent tool for hypergraph analysis. 
}

\begin{figure}[h!]
    \centering
    \includegraphics[height=225pt]{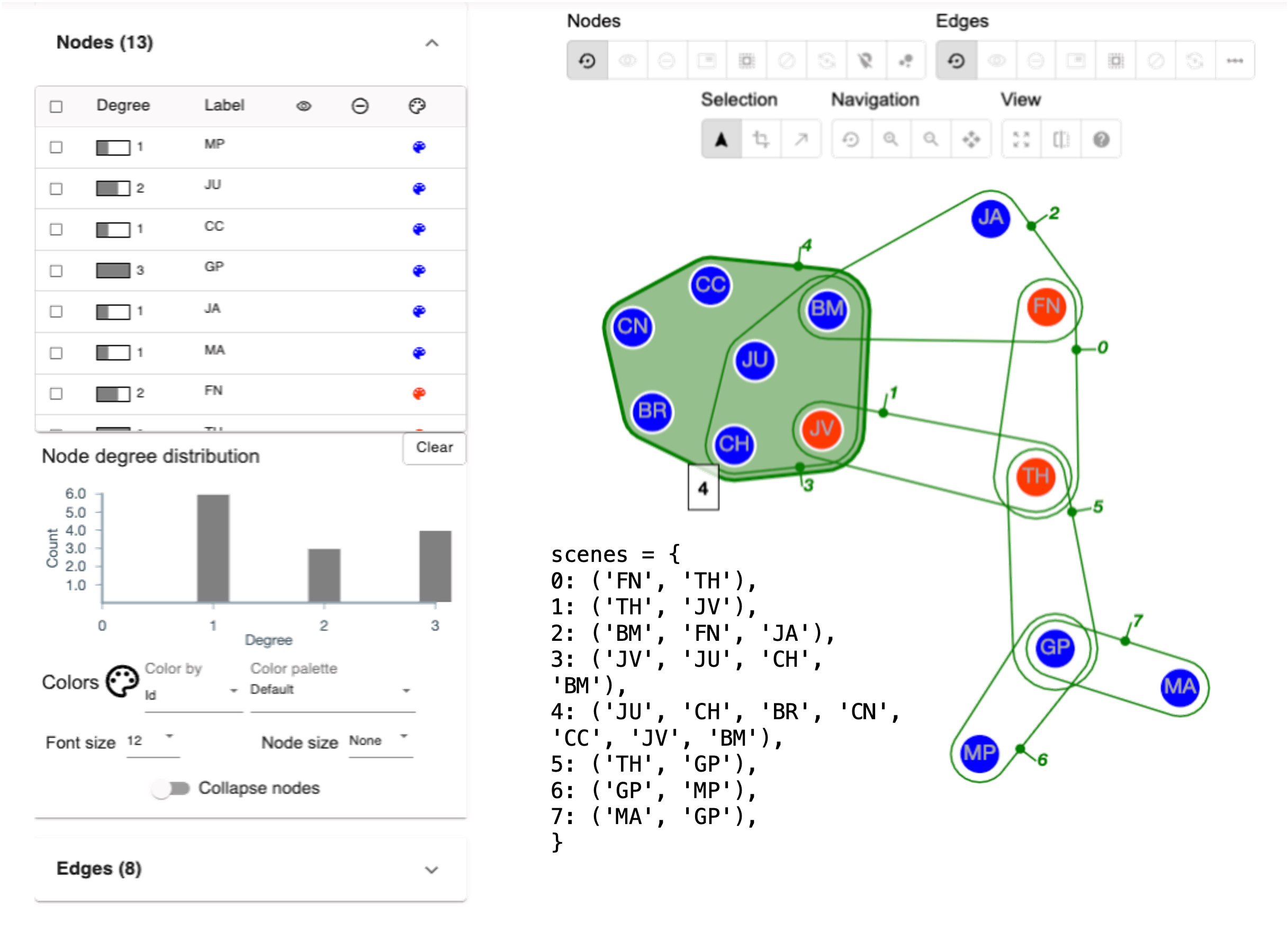}
    \caption{HNX-Widget visualization of a Scene to Character mapping from the LesMis dataset \cite{knuth1993}.}
    \label{fig:hnxexample}
    \end{figure}

\section{Statement of need}
For more than a century, graph theory has provided powerful methods for studying network relationships among abstract entities.
Since the early 2000's, software packages such as NetworkX \cite{hagberg2020} and Igraph \cite{csardi2006} have
made these theoretical tools available to data scientists for studying large data sets.
Graphs represent pairwise interactions between entities, but for many network datasets this is a severe limitation.
In 1973, hypergraphs were introduced by Claude Berge \cite{Berge1973Graphs} as a strict generalization of graphs: a hyperedge in a hypergraph can contain any number of nodes, including 1, 2, or more.
Hypergraphs have been used to model complex network datasets in
areas such as the biological sciences, information systems, the power grid, and cyber security.
Hypergraphs strictly generalize graphs (all graphs are (2-uniform) hypergraphs), and thus can represent additional data complexity and have more mathematical properties to exploit (for example, hyperedges can be contained in other hyperedges). As mathematical set systems, simplicial and homological methods from
Algebraic Topology are well suited to aid in their analysis \cite{Joslyn2021,Torres2021}.
With the development of hypergraph modeling methods, new software was required to support
experimentation and exploration, which prompted the development of HyperNetX.

\section{Related Software}
Due to the diversity of hypergraph modeling applications, hypergraph software libraries are 
often bootstrapped using data structures and methods most appropriate to their usage. 
A nice compendium of many of the hypergraph libraries created in the last decade can be found in \cite{Kurte2021}.
HNX leads the effort to share library capabilities by specifying a Hypergraph Interchange Format (HIF) 
for storing hypergraph data as a JSON object. Since hypergraphs can store metadata on its nodes, 
hyperedges, and incidence pairs, a standardized format makes it easy to share hypergraphs across libraries.

\begin{figure}[h!]
    \centering
    \includegraphics[scale=.30]{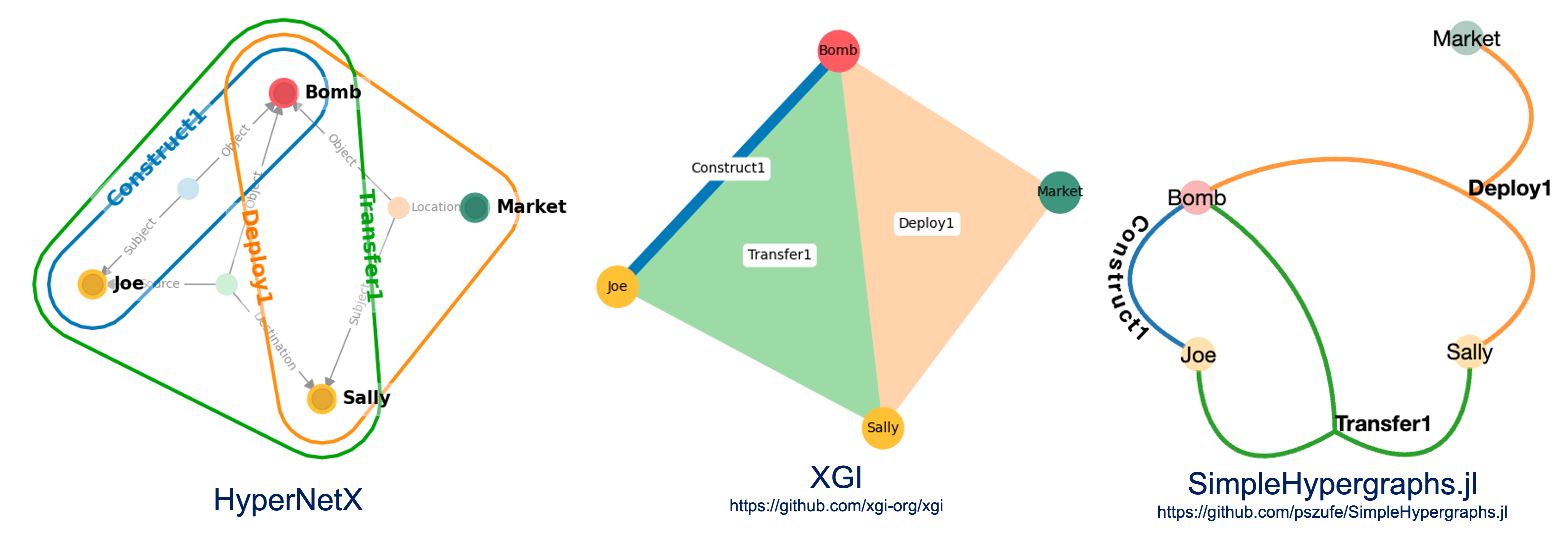}
    \caption{Visualizations from hypergraph libraries based on the bipartite graph seen in grey
    under the HyperNetX visualization (left side): XGI (Center), \cite{Landry2023} and SimpleHypergraph (Right), \cite{Szufel2019}.}
    \label{fig:3graphs}
    \end{figure}
  
\section{Overview of HNX}
HNX serves as a platform for the collaboration and sharing of hypergraph
methods within the research community.
Originally intended to generalize many of the methods from NetworkX
to hypergraphs, HNX now has implementations for many hypergraph-specific metrics.
While graph paths can be measured by length,
hypergraph paths also have a width parameter \textit{s}, given by the minimum intersection size
of incident hyperedges in the path \cite{Aksoy2020Hypernetwork}.
HNX uses this \textit{s} parameter in many of
its core methods as well as in its \textit{s-}centrality module.
As set systems, hypergraphs can be viewed as subsets of abstract simplicial
complexes – combinatorial projections of geometric objects constructed from points, line
segments, triangles, tetrahedrons, and their higher dimensional analogues.
HNX's Simplicial Homology module identifies and computes the \textit{voids} of different dimensions
in the simplicial complexes generated by modestly sized hypergraphs.
These objects, which are used for defining the \textit{Homology Groups}
studied by Algebraic Topologists, offer new metrics for exploratory
data science.
 
As a collaborative platform, HNX contains contributed modules
and tutorials in the form of Jupyter notebooks
for Laplacian clustering, clustering and modularity, synthetic
generation of hypergraphs, and Contagion Theory.
In its latest release, HNX 2.0 uses Pandas dataframes as its underlying data structure,
making the nodes and hyperedges of a hypergraph as accessible as the
cells in a dataframe.
This simple design allows HNX to import data from semantically
loaded graphs such as property graphs and knowledge graphs,
in order to model and explore their higher order relationships.
Because it is open source, HNX provides a unique opportunity for
hypergraph researchers to implement their own methods built from
HNX and contribute them as modules and Jupyter tutorials to the HNX user community.

\section{HNX-Widget: Interactive Diagrams of Hypergraphs}
HNX interfaces with an add-on widget for the Jupyter Notebook
computational environment, enabling users to view and interactively
explore hypergraphs, see Figure 1.
The main features of the tool are: 1) adjustable layout 2) advanced
selection and 3) visual encoding of node and edge properties.
 
The primary tool of the widget is an Euler diagram visualization approach that shows nodes as circles and hyperedges as closed curves surrounding the nodes/circles they contain. The tool uses a force directed layout of the underlying bipartite representation to determine the position of the nodes. Hyperedges are then drawn as convex hulls given the node positions. To remove overlapping hyperedges the user and can move and pin nodes as desired.
Selection of nodes and hyperedges is available via clicking, brushing, and advanced methods. For example, the set of selected nodes can be expanded by selecting all nodes contained by selected edges, and vice versa.

 

Metadata may be attached to the tool by providing tabular data via two optional data frames indexed by node and hyperedge identifiers. Single rows from these tables become visible to the user in a tooltip when they hover over a node or hyperedge in this tool. The user can configure the tool to encode the node size or color and the hyperedge color using a column in these data tables.
 
\section{Projects using HNX}
HNX was created by the Pacific Northwest National Laboratory. It has provided data analysis and visualization support for academic papers in subject areas such as biological systems \cite{Feng2021,Colby2023}, cyber security \cite{Joslyn2020DNS}, information systems \cite{Molnar2022Application}, neural networks \cite{Praggastis2022SVD}, knowledge graphs \cite{joslyn2018}, and the foundations of hypergraph theory \cite{Vazquez2022Growth}.
 
\bibliographystyle{plain}
\bibliography{paper}
\end{document}